# CVE based classification of vulnerable IoT systems

Grzegorz J. Blinowski[1] [0000-0002-0869-2828] and Paweł Piotrowski[2]

[1,2] Institute of Computer Science, Warsaw University of Technology, Nowowiejska 15/19, 00-665 Warszawa, Poland
g.blinowski@ii.pw.edu.pl

**Abstract.** Common Vulnerabilities and Exposures database (CVE) is one of the largest publicly available source of software and hardware vulnerability data and reports. In this work we analyze the CVE database in the context of IoT device and system vulnerabilities. We introduce a real-world based classification of IoT systems. Then, we employ a SVM algorithm on selected subset of CVE database to classify "new" vulnerability records in this framework. The subset of interest consists of records that describe vulnerabilities of potential IoT devices of different applications, such as: home, industry, mobile controllers, networking, etc. The purpose of the classification is to develop and test an automatic system for recognition of vulnerable IoT devices and to test completes, sufficiency and reliability of CVE data in this respect.

## 1 Introduction and Background

### 1.1 IoT Architecture – Outline

IoT can be most broadly defined as an interconnection of various uniquely addressable objects through communication protocols. Narrowing down the above said, we can describe it as a communication system paradigm in which the objects of everyday life, equipped with microcontrollers, network transmitters, and suitable protocol stacks that allow them communicate with one another and, via ubiquitous cloud infrastructure and also with users, become an integral part of the Internet environment [1].

Here, we will consider an IoT model consisting of three major levels:

- **Perception and execution layer**, which encompasses a wide range of "smart" devices ranging from RFID and NFC enabled tags, environmental sensors and actuators, various home appliances, mobile terminals, smart phones, etc. This also includes a wide range of industry SCADA devices and smart vehicle components, ubiquitous in today's vehicles (cars, trucks, etc.). A separate role must be attributed to controllers or managing stations used both in home and industrial application. A controller is simply a PC, a tablet or a mobile phone used to manage local IoT infrastructure, and to visualize data.

- **Network layer** which provides heterogeneous communication infrastructure based on multiple network standards such as: s Wi-Fi, 3G/LTE, Z-wave, ZigBee, 6LoWPAN, VLC, mIP [2] and Ethernet together with the standard internet protocol suite (IPv4 / IPv6 and transport layer UDP/TCP stack).
- **Cloud or application layer** which integrates, manages and analyzes data from IoT devices. The cloud not only gathers data and manages the "things" and "core" layer, but acts as a ubiquitous service provider for end-users, according to the Service Oriented Approach (SOA) paradigm.

- The above IoT model is compatible with the reference architecture model proposed by the EU FP7 IoT-A project [3] and the IoT-A tree structure [4].

### 1.2 IoT Applications

IoT is widely, although mostly anecdotally, known as a network of household appliances – from PC equipment and peripherals to fridges, coffee machines, etc. However, the scope of IoT deployments is much wider, and covers the following areas [1,5,6,7]:

- Smart Cities – building structural health monitoring, noise mapping, traffic congestion monitoring and "smart roads"; smart lighting; waste management.
- Smart environment – weather monitoring; disaster early warning systems, e.g. flood detection and volcano monitoring); water quality monitoring; chemical leakage and pollution level detection.
- Smart agriculture and farming – fertilizer, pesticide and irrigation monitoring, crop level monitoring; hydroponic plant monitoring and control; animal tracking.
- Smart Grid – electrical energy consumption monitoring and management.
- Manufacturing – this covers a wide range of industry process control systems – mechanical, chemical, etc. This range of applications is often referred to as IToT (Industrial IoT); the systems themselves are referred to as SCAD (Supervisory Control and Data Acquisition).
- Industrial Security and sensing – gas level and leakage detection in industrial environments, radiation level measurement.
- eHealth – patient surveillance and assistance in medical facilities.
- Home automation ("Smart homes") – energy and water use monitoring, remotely controlled appliances, door cameras, locks, alarms, etc.

### 1.3 Security issues with IoT systems

Security issues with IoT environments have been widely discussed and publicized. In some cases, when the compromised system was widely used, for example as commodity type home appliance, or when the effects of security exploitation where widely conspicuous (e.g. in the case of Mirai botnet [8]) the awareness of IoT insecurity has even reached general public.

We can distinguish two general kinds of IoT threats: 1. **threats against IoT** and 2. **threats from IoT**. 1 Threats against IoT occur when a flaw in an IoT device or application, on the perception, network or cloud level is exploited by the hacker, and the device or application is compromised - i.e. a full or limited access to its functions and data is gained by an attacker. 2 In case of threats from IoT, the compromised infrastructure is used to conduct various attacks against IoT or Internet-connected devices. Again, the Mirai botnet can serve as an example - when a multitude of compromised webcams and other devices were used to conduct a massive DDoS attack.

In [9] the authors have proposed five "dimensions" relating to IoT security: hardware, operating system/firmware, software, networking and data:

- **Hardware security** is critical when an attacker can physically access the device. Through the hardware backdoors, software level integrity checking can be bypassed by disabling the checking functionality or booting via forged firmware. Almost all IoT devices have hardware vulnerabilities which may be exploited (see a reference database - [10]). The vulnerability list mechanisms include, but are not limited to: debugging ports, multiple boot options, and unencrypted flash memory [11]–[13]. Microcontrollers (MCUs) which are broadly used in industry applications (SCADA) as well as in automobiles and home automation are also prone to hardware level vulnerabilities. This includes for example attacks against EEPROM contents via JTAG/SPI ports, clock glitch attacks which have been shown to compromise AES encryption via fault injection techniques [14]
- **Operating system, firmware and software security and privacy** - relates to all three IoT layers: perception, network and cloud. Software security issues are similar to those in the traditional computer systems. Trustworthy operating systems should be used at the perception layer to reduce the risk of remote compromise. However, in practice, this is rarely the case. The controller application is often installed on a PC or a smartphone and software secure measures should be applied in order to prevent the attack against it. The cloud layer security also cannot be blindly trusted, for example: servers installed on Amazon EC2 are secured from the cloud provider's point of view, but not from the point of view of installed application and have to be secured by whoever deploys the servers. Other software level security risks specific to the IoT environments described recently include for example: public and private SSL key pairs discovered by static analysis on a large number of unpacked firmwares; a large-scale automated dynamic analysis using the Metasploit Framework of various firmwares was conducted and a large number potential exploits were discovered [15]; a buffer overflow exploit (which can be used to execute any code on the device) was found by analyzing Home Network Administration Protocol (HNAP) [16]; stack-based buffer overflow of the general library glibc [17] was exploited to attack several home hubs [18].
- **Network Security and Privacy** - as a networked system a whole IoT environment has to be secured from end to end. Encryption and authentication should be used consistently, but often is not. Two functions specific to home IoT devices are pairing and binding. In the pairing process the controller needs to connect to the IoT device in order to configure the "thing". Most SOHO IoT devices allow any controller in proximity to conduct pairing with no additional security measures. This may be acceptable in home environment, but in a large-scale deployment in a public environment anybody with access to the devices can reconfigure and break into the system. The binding process starts after pairing succeeds and establishes necessary access credentials for the thing in order to control it. Week passwords are one of the typical security issues in this phase. Many manufacturers fail to provide necessary protection for their networked IoT devices, the above quoted Mirai attack being just one of the examples of weak password exploitation. Many attacks relating to binding have been analyzed and described, for example for: wearable devices [19], surveillance camera systems [20], Phillips Hue light bulb system [21], etc.

- **Cloud and data** - the cloud collects data from the perception layer, and is responsible for maintaining proper data security. The cloud often handles authentication and associated services and is a peer in end-to-end encryption of transmitted data. Application compromised on the cloud level exposes a significant amount or perhaps the whole of the collected data. However, big data collected by the cloud can also help to increase security. For example, it can be used to distinguish between legitimate and illegal usage patterns, it can also prevent (at least to some extent) DDoS attacks.

To summarize this section: the majority of security problems emerging in today's IoT systems result directly from buggy, incomplete or outdated software and hardware implementations. A major protocol flaw design error (such as Heartbleed [22] and DROWN [23]) are much rarer. As can easily be verified in public domain vulnerability databases, the number of products reported with serious security flaws is growing year by year.

### 1.4 Scope of this work and related research

In this work, we propose a classification of device-related (i.e. not "pure software") vulnerability data for IoT and IIoT equipment. We have divided the CVE records from a public database into 7 distinct categories (e.g.: home equipment, SCADA devices, network infrastructure systems, etc.). The database samples were hand-classified by us based on the expert knowledge. We then used support vector machine (SVM) classifier on the device and vulnerability data to predict categories of "new" vulnerabilities – for example data from year 2017 was used to classify 2018's data, etc. The purpose was to predict, and (if possible) prevent and mitigate threats resulting from new vulnerabilities. This is a difficult task given the size of the database and the rate of its growth – each day tens of new records are added to the CVE database alone. Hence, when a new vulnerability or exploit is discovered it is often critical to learn its scope by automatics means, as fast as possible.

There has been some prior research on automatic analysis and classification of vulnerability databases: In [24,25] models and methodologies of categorizing vulnerabilities from CVE database according to their security types based on Bayesian networks. In [26] Topic Models were used to analyze security trends in CVE database with no prior (expert) knowledge. Huang et. al. [27] proposed recently an automatic classification of records from NVD database based on deep Neural Network, the authors compared their model to Bayes and KNN models and found it superior. All of the above cited research was focused on categorizing software aspect of vulnerabilities, with categories such as for example: SQL injection, race condition, cryptographic errors, command injection, etc. According to our knowledge no prior work was done regarding categorizing of the impacted equipment: system or device – our work tries to address this gap.

This paper is organized as follows: in section 2 we describe the contents and structure of the CVE database; we also describe related: CPE (Common Platform Enumeration) and NVD (Network Vulnerability Data) records. In section 3 we introduce our proposed classes of IoT devices; we discuss briefly SVM classifier methods and the measures we used to test classifiers quality. In section 4 we present the results of the classification. Our work is summarized in section 5.

# 2 Structure and contents of CVE Database

## 2.1 The Common Vulnerability and Exposures (CVE) database

The Common Vulnerability and Exposures (CVE) database hosted at MITRE is one of the largest publicly available source of vulnerability information. As the CVE's FAQ [28] states: "C*VE is a list of information security vulnerabilities and exposures that aims to provide common names for publicly known problems. The goal of CVE is to make it easier to share data across separate vulnerability capabilities (tools, repositories, and services) with this 'common enumeration." CVE assigns identifiers (CVE-IDs) to publicly known It product vulnerabilities.* Across organizations, IT-security solutions vendors, and security experts, CVE has become the de facto standard of sharing information on known vulnerabilities and exposures.

In this work we use an annotated version of the CVE database, known as National Vulnerability Database (NVD) which is hosted by National Institute of Standards and Technology (NIST). NVD is created on the basis of information provided by MITRE (and through the public CVE site). NIST adds other information such as structured product names and versions, and also maps the entries to CWE names. NVD feed is provided both in XML and JSON formats structured in year-by-year files, as a single whole-database file and as an incremental feed reflecting the current year's vulnerabilities.

```
<?xml version='1.0' encoding='UTF-8'?>
  <nvd xmlns:scap-core="http://scap.nist.gov/schema/scap-core/0.1"
xmlns:cvss="http://scap.nist.gov/schema/cvss-v2/0.2"
xmlns:vuln="http://scap.nist.gov/schema/vulnerability/0.4"
xmlns:xsi="http://www.w3.org/2001/XMLSchema-instance" ...>
   <entry id="CVE-2017-3741">
      <vuln:vulnerable-configuration id="http://nvd.nist.gov/">
        <cpe-lang:logical-test operator="OR" negate="false">
            <cpe-lang:fact-ref name=
              "cpe:/a:lenovo:power_management:1.67.12.19"/>
            <cpe-lang:fact-ref name=
              "cpe:/a:lenovo:power_management:1.67.12.23"/>
          </cpe-lang:logical-test>
      </vuln:vulnerable-configuration>
      <vuln:vulnerable-software-list>
        <vuln:product>
          cpe:/a:lenovo:power_management:1.67.12.19</vuln:product>
      </vuln:vulnerable-software-list>
      <vuln:cve-id>CVE-2017-3741</vuln:cve-id>
      <vuln:published-datetime>2017-06-04T17:29:00.387-
         04:00</vuln:published-datetime>
      <vuln:last-modified-datetime>2017-06-13T13:13:17.
         827-04:00</vuln:last-modified-datetime>
      <vuln:cvss>
        <cvss:base_metrics>
          <cvss:score>2.1</cvss:score>
          <cvss:authentication>NONE</cvss:authentication>
          <cvss:confidentiality-impact>NONE
           </cvss:confidentiality-impact>
          <cvss:integrity-impact>PARTIAL</cvss:integrity-impact>
          <cvss:availability-impact>NONE</cvss:availability-impact>
        </cvss:base_metrics>
      </vuln:cvss>
      <vuln:cwe id="CWE-254"/>
      <vuln:references xml:lang="en" reference_type="VENDOR_ADVISORY">
        <vuln:source>CONFIRM</vuln:source>
        <vuln:reference href=
          "https://support.lenovo.com/us/en/product_security/
           LEN-14440" xml:lang="en">https:...//.../LEN-14440
        </vuln:reference>
      </vuln:references>
      <vuln:summary>In the Lenovo Power Management driver before
         1.67.12.24, a local user may alter ... This
         issue only affects ThinkPad X1 ... generation.</vuln:summary>
```

```
  </entry>
  </nvd>
```

**Fig. 1.** A single simplified NVD record from NIST CVE feed (some less relevant fields have been abbreviated or omitted).

**Fig. 1** contains a sample (simplified) record from the NVD database. Fields which are relevant for further discussion are as follows:

- `entry` contains record id as issued by MITRE, the id is in the form: CVE-yyyy-nnnnn (eg. CVE-2017-3741) and is commonly used in various other databases, documents, etc to refer to a given vulnerability
- `vuln:vulnerable-configuration` and `vuln:vulnerable-software-list` identifies software and hardware products affected by a vulnerability. This record contains the description of a product and follows the specifications of the Common Platform Enumeration (CPE) standard. Because the vulnerability's scope may be complex – for example it may refer to a particular version of software on a given hardware platform, the product description is formatted as a structured, logical, AND-OR expression.
- `cpe-lang` - The basic record of vuln structure More information on CPE format will be provided in the next section.
- `vuln:cvs` and `cvss:base_metrics` describe the scope and impact of the vulnerability. This data allows to identify real-world consequences of the vulnerability, that is its access-, availability- and confidentiality impacts. For example it reports if the bug allows for remote system takeover, is it a data breach, etc.
- `vuln:cwe` contains a reference to is a community-developed list of common software security weaknesses (CWE) [29] database. CWE is hosted by MITRE and contains formal list of software weakness types. In simplest terms CWE ID identifies type of bug that caused the vulnerability.
- `vuln:references` may contain URL providing additional information regarding the vulnerability.
- `vuln:summary` holds a vulnerabilities short informal description.

### 2.2 Common Platform Enumeration (CPE)

CPE is a formal naming scheme for identifying: applications, hardware devices, and operating systems. CPE is part of the Security Content Automation Protocol (SCAP) 5 standard [30], which was proposed by the National Institute of Standards and Technology (NIST). Here we will refer to the most recent version 2.3 of CPE. The CPE naming scheme is based on a set of attributes called Well-Formed CPE Name (WFN) [31]. The following attributes are part of this format: part, vendor, product, version, update, edition, language, software edition, target software, target hardware, and other (not all attributes are always present in the CPE record, very often "update", and the following ones are omitted from the record). Currently, CPE supports two formats: URI (defined in CPE version 2.2) and formatted string (defined in CPE version 2.3). The CVE database uses URI format and we will only discuss this format further on.

For example: in the following CPE record **`cpe:/h:moxa:edr-g903:-`** the attributes are as follows: **`part:h`** (indicating hardware device), **`vendor:moxa`**, **`product:edr-903`**, version, and the following attributes are not provided. As a second example let us consider a complex `vuln:vulnerable-configuration` record – **Fig. 2** below.

```
<vuln:vulnerable-configuration id="http://nvd.nist.gov/">
  <cpe-lang:logical-test operator="AND" negate="false">
    <cpe-lang:logical-test operator="OR" negate="false">
     <cpe-lang:fact-ref name=
       "cpe:/o:d-link:dgs-1100_firmware:1.01.018"/>
     </cpe-lang:logical-test>
     <cpe-lang:logical-test operator="OR" negate="false">
     <cpe-lang:fact-ref name="cpe:/h:d-link:dgs-1100-05:-"/>
     <cpe-lang:fact-ref name="cpe:/h:d-link:dgs-1100-05pd:-"/>
     <cpe-lang:fact-ref name="cpe:/h:d-link:dgs-1100-08:-"/>
     <cpe-lang:fact-ref name="cpe:/h:d-link:dgs-1100-08p:-"/>
     <cpe-lang:fact-ref name="cpe:/h:d-link:dgs-1100-10mp:-"/>
      ...
    </cpe-lang:logical-test>
  </cpe-lang:logical-test>
</vuln:vulnerable-configuration>
```

**Fig. 2.** A vulnerable configuration record from CVE – a logical expression build from CPE identifiers.

The record shown on **Fig. 3** refers to particular operating systems version (firmware), namely: `cpe:/o:d-link:dgs-1100_firmware:1.01.018` on a list of distinct hardware devices: `cpe:/h:d-link:dgs-1100-05:-`, `cpe:/h:d-link:dgs-1100-05pd:-`, etc. CPE does not identify unique instantiations of products on systems, rather, it identifies abstract classes of products. The first component of the CPE descriptor is "part" it can take the following values: a – for application, h – for hardware, o – for operating system.

CPE records are maintained in a separate database "CPE Dictionary" [31] which is distributed in XML format, it contains product records with URL reference to product description – **Fig. 1Fig. 3**.

```
<cpe-item name="cpe:/h:d-link:dap-1320:-">
  <title xml:lang="en-US">D-Link DAP-1320</title>
    <references>
      <reference href="http://us.dlink.com/products/access-points-
range-extenders-and-bridges/wireless-range-extender/">Vendor web-
site</reference>
    </references>
    <cpe-23:cpe23-item name="cpe:2.3:h:d-link:dap-1320:-
:*:*:*:*:*:*:*"/>
</cpe-item>
```

**Fig. 3.** A CPE record defining a named device.

### 2.3 Discussion

The NVD database is distributed as XML and JSON feeds, it is also possible to download the whole historical data package (starting from 1999, but records compliant with the current specification are available for data generated since 2002). In addition, there is also an on-line search interface. The database, as at the beginning of 2020 contains over 120 000 records in total, and on the average the number of records increases year by year. Due to historical reasons – a long time-span of data collection it neither completely consistent, nor free of errors. Older records lack some information (for example vulnerability score); URL in the reference field may be outdated; there are approximately 900 records without CPE identifier; there exists a large number of records with CPEs inconsistent or not present in the CPE dictionary (approx. 100 000 CPEs). In general, the binding between the vulnerability description and the product concerned may be problematic, because it is only provided by the CPE – for example there is no “plain language” product description, or classification present in the database. Product names containing non-ASCII or non-European characters also pose a problem, as they are recoded to ASCII often inconsistently or erroneously.

Lack of record classification on the CVE or CPE level (except for the “application, OS or hardware” attribute in the CPE) is especially cumbersome, because there is no easy or obvious way to differentiate products. Essentially, it is impossible to extract data relating to, for example: web servers, home routers, IoT home appliances, security cameras, cars, SCADA systems, etc., without a priori knowledge of products and vendors.

## 3 CVE Data classification and analysis

### 3.1 Data selection

For the classification purposes we have selected only records with the CPE “part” attribute set to “h” (hardware records), namely, the selection criteria was: if any of the records in `vuln:vulnerable-configuration` section contains CPE with part = =”h” then the record was selected for further consideration. Other records were discarded. The reason is the following: all or most of the “hardware” type records refer to devices or systems which can potentially be a component of the perception or network layer of the IoT or IIoT architecture. We have also narrowed down the timeframe to data from years 2010-2019 (data from first quarter of 2019 was taken into account). **Fig. 4** shows the number of all records from this time period.

Hand analysis of selected vulnerability data led us to a grouping of records into 7 distinct classes as follows:

- **H** – Home and SOHO devices; routers, on-line cameras and monitoring, other customer grade-appliances.
- **S** – SCADA and industrial systems, automation, sensor systems, non-home IoT appliances, car and vehicles (subsystems), medical devices, industrial video recorders and surveillance systems.

- **E** – Enterprise, Service Provider (SP) hardware (routers, switches, enterprise Wi-Fi and networking) – this constitutes mainly the network level of IoT infrastructure.
- **M** – mobile phones, tablets, smart watches and portable devices - this constitutes the "controllers" of IoT systems.
- **P** – PCs, laptops, PC-like computing appliances and PC servers (enterprise) – this constitutes the "controllers" of IoT systems.
- **A** – other, non-home appliances: enterprise printers and printing systems, copy machines, non-customer storage and multimedia appliances.

The reason for the above classification was practical – the key distinction for an IoT component in reference to its security vulnerability is the market and scope of its application (home use, industrial use, network layer, etc.). On the other hand we are limited by the description of the available data – it would impossible to use a finer-grain classification. Also, it is not practical to introduce to many classes with small number of members, because the automatic classification quality suffers in such case (As we have in fact learned in same cases). Sample devices from NVD database are shown in Table 1.

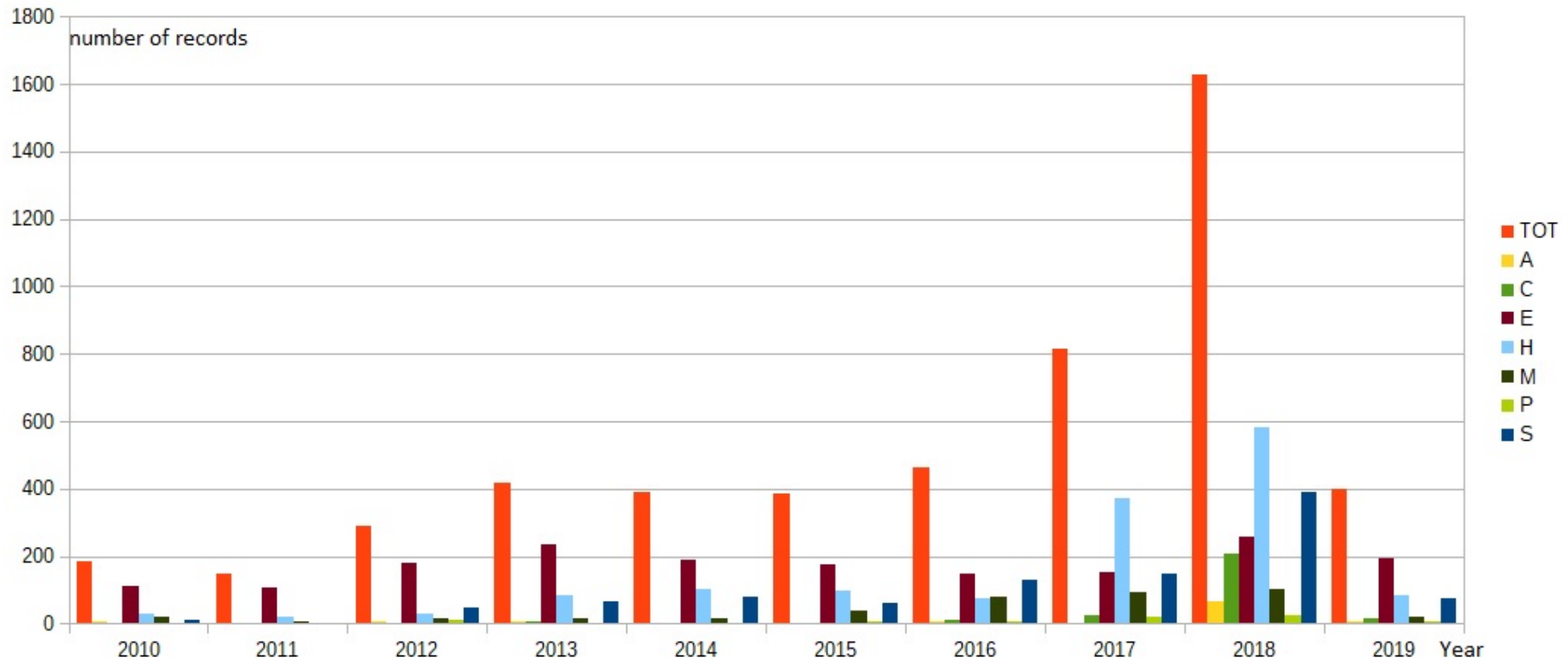

**Fig. 4.** Number of "h" records classified in the time range of 2010-2019. TOT is the total number of records, other bars refer to assigned classes. Only the first quarter of 2019 was considered, hence a smaller record number.

**Table 1.** Devices from NVD database – samples from each of the proposed classes.

| Class | Sample systems (vendor, name) | Notes |
|---|---|---|
| **H** | • D-link Dir-815<br>• Opticam i5<br>• Meetcircle Circle With Disney<br>• Amazon Amazon_key | • Network & Wi-Fi router<br>• Home surveillance camera<br>• Parental control device<br>• Home Access Device |
| **S** | • Siemens Sinumerik 828d<br>• Yokogawa FCI<br>• Mbusa cockpit<br>• Vivotek camera | • industrial machine tool controller<br>• autonomous controller<br>• car cockpit automation<br>• surveillance camera |
| **E** | • Juniper SRX100 SRX110<br>• CISCO staros asr_5000<br>• Citrix Netscaler Gateway<br>• Polycom QDX6000 | • Family of network firewalls<br>• Router & access device<br>• Network load balancing system<br>• Video-conferencing system |
| **M** | • Samsung Galaxy S6<br>• Amazon Kindle Fire<br>• Mi mi router 3<br>• Huawei Watch 2 | • Smart phone<br>• Tablet / e-book reader<br>• Portable Wi-Fi/LTE router<br>• Smartwatch device |
| **P** | • HP integrated lights-out<br>• HP nonstop_server<br>• Intel s7200ap | • Server management module<br>• Server platform<br>• Server main board |
| **A** | • Drobo 5n2<br>• TBK Vision tbk-dvr4216<br>• Ricoh d2200 | • Enterprise data storage system<br>• Enterprise DVR system<br>• Enterprise Printing system |

### 3.2 Data analysis methodology

We build classifiers by training linear support vector machines (SVM) [32] on the features of “hardware” vulnerability records extracted from the NVD database. The feature vector contains:

- Vendor name,
- product name and other product data from CPE (if supplied),
- vulnerability description,
- error code (CWE).

The steps of the process of building a classifier are the following:

- Preprocessing of input data (removal of stop-words, lemmatization, etc.),
- feature extraction, i.e. conversion of text data to vector space,

- training of the linear SVM.

We use a standard linear SVM, which computes the maximum margin hyperplane that separates the positive and negative examples in feature space. The classification based on a linear SVM generates the hyperplane that maximizes the distance of the most borderline training examples to the linear decision plane (or boundary). Alternative methods include: k-nearest neighbor, Bayesian classifiers and Neural Nets. We have conducted some experiments with Neural Nets, but finally decided to use SVM, as it proved itself to be: fast, efficient and well suited for text-data classification. With SVM method the decision boundary is not only uniquely specified, but statistical learning theory shows that it yields lower expected error rates when used to classify previously unseen examples [32,33] - i.e. it gives good results when classifying new data.

We have used Python 3.7.1 with NLTK 3.4.1 [34] and scikit-learn 0.21.3 [35] libraries. NLTK was used to pre-process the text data, while scikit contains SVM algorithms together with tools for computing the classification quality metrics.

### 3.3 Classification measures

To benchmark classification result we use two standard measures: precision and recall. We define *precision* (eq. (1)) as a fraction of relevant instances among the retrieved instances; we define *recall* (eq. (2)) as the fraction of the total amount of relevant instances that were actually retrieved. In other words: precision indicates the ratio of true positives to the sum of true positives and false positives, while recall is calculated as the ratio of true positives the sum of true positives and false negatives (elements belonging to the current category but not classified as such). Finally, as a concise measure of we will use the F1 score, also known as balanced F-score. The F1 score can be interpreted as a weighted average of the precision and recall, where an F1 score reaches its best value at 1 and worst score at 0. The relative contribution of precision and recall to the F1 score are equal. The formula for the *F1* score is given in eq. (3)

$$precission = {}^{TP}/_{(TP+FP)} \quad (1)$$

$$recall = {}^{TP}/_{(TP+FN)} \quad (2)$$

$$F1 = 2 * \frac{precision * recall}{(prceision + recall)} \quad (3)$$

## 4 Classification results

### 4.1 Data selection and classification

We have tested the classifier for historical data in one year intervals. For example, to classify data from 2018 we have used records from the following ranges: 2014-2017, 2015-2017, 2016-2017 and 2017, etc. On **Fig. 5** we show confusion matrices trained on data ranging from 2014 to 2017 used to classify data for 2018. From a good classifier we would expect a majority of records on the diagonal. Here the classification is not perfect, for example - for training data from year range 2014-2017: 109 H type records were marked as E and 108 as S class; only 62% were correctly classified (recall). When only data from 2017 was used, perhaps surprisingly, the classification is more accurate: 489 / 85% records of the H type were labelled correctly (recall); however for S and E classes only 55% were correctly identified. For classes with a low number of records (C, M, P) the classification falls below 50%. On **Fig. 6.** Precission and recall for 2018 records based on training data from: 2014-2017, 2015-2017, 2016-2017 and 2017 (category "A" was removed).

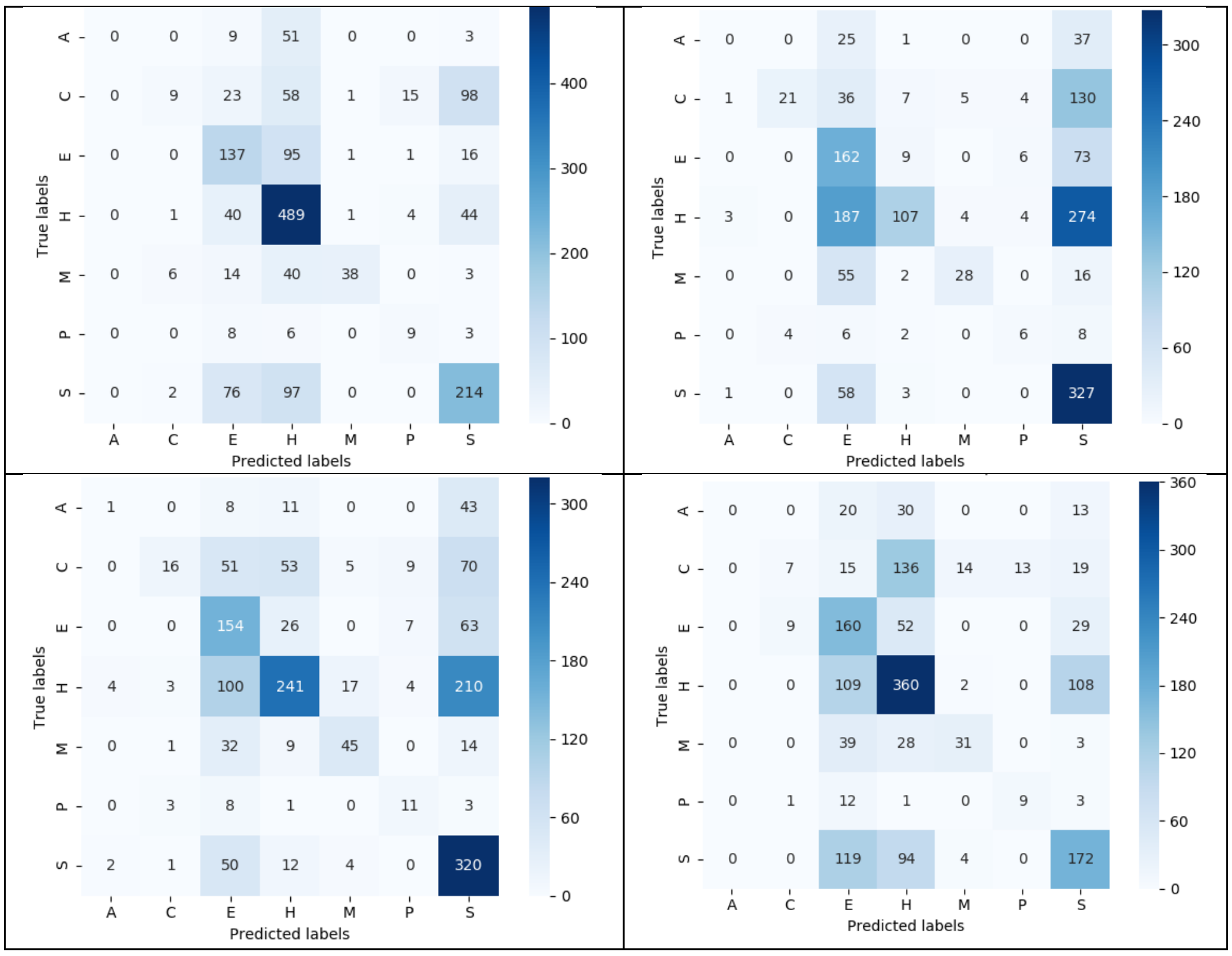

**Fig. 5.** Classification of records from 2018 based on training data from: 2017, 2016-2017, 2015-2017 and 2014-2017 (left to right, top-bottom). Numbers show the number of records.

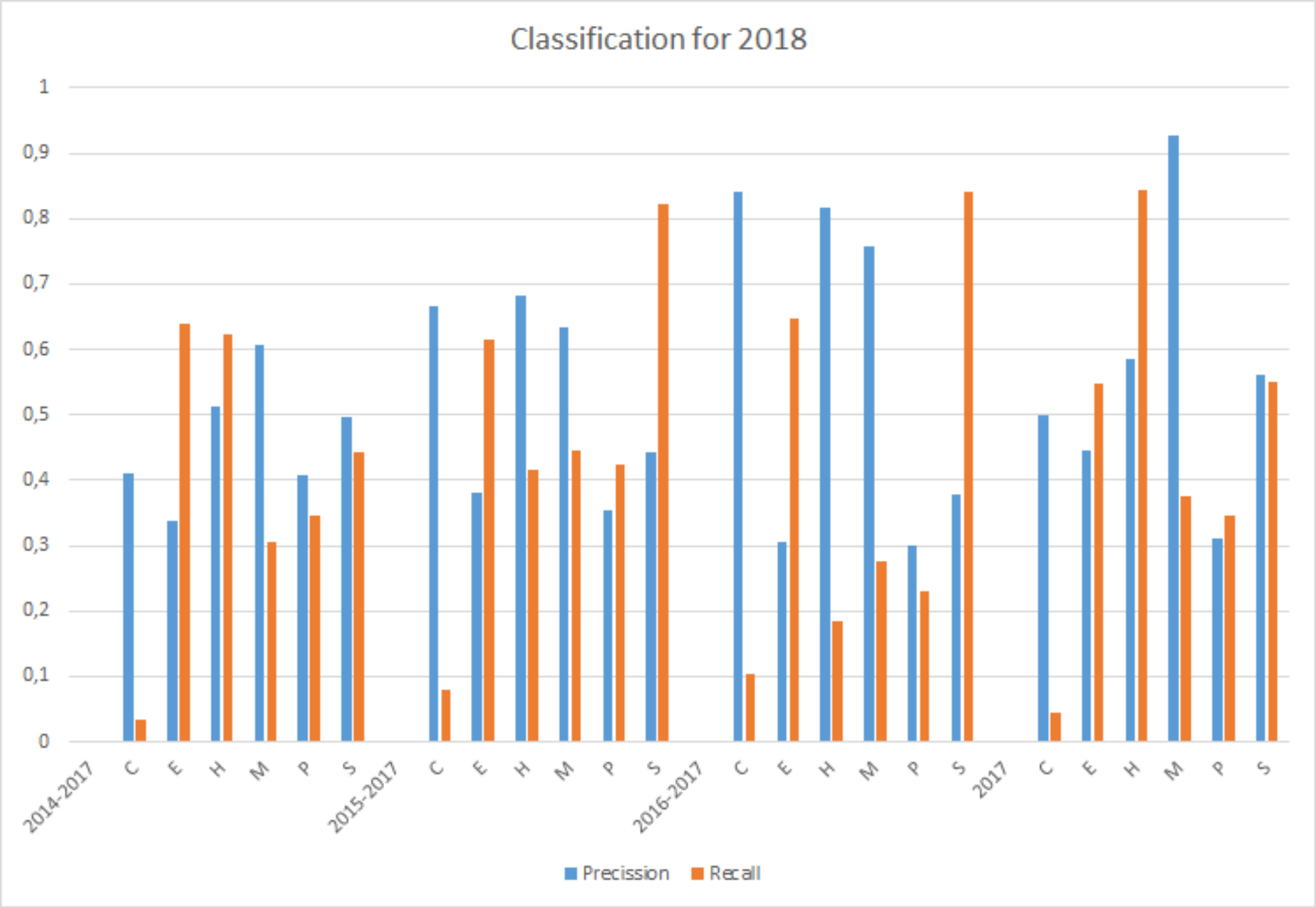

**Fig. 6.** Precission and recall for 2018 records based on training data from: 2014-2017, 2015-2017, 2016-2017 and 2017 (category "A" was removed).

Similar trends are visible for classified data from the first quarter of 2019 based on SVM trained on records from 2015-2018 range – for E, H and S classes precision is within the range of 70%-90% (with an exception of H class where it is only 44%), and recall falls in the similar range of 70%-90%.

On **Fig. 8** we have shown result for years 2015 and 2016 – for major classes the classification for years ranging from 2011-214 and 2017 also show similar trends.

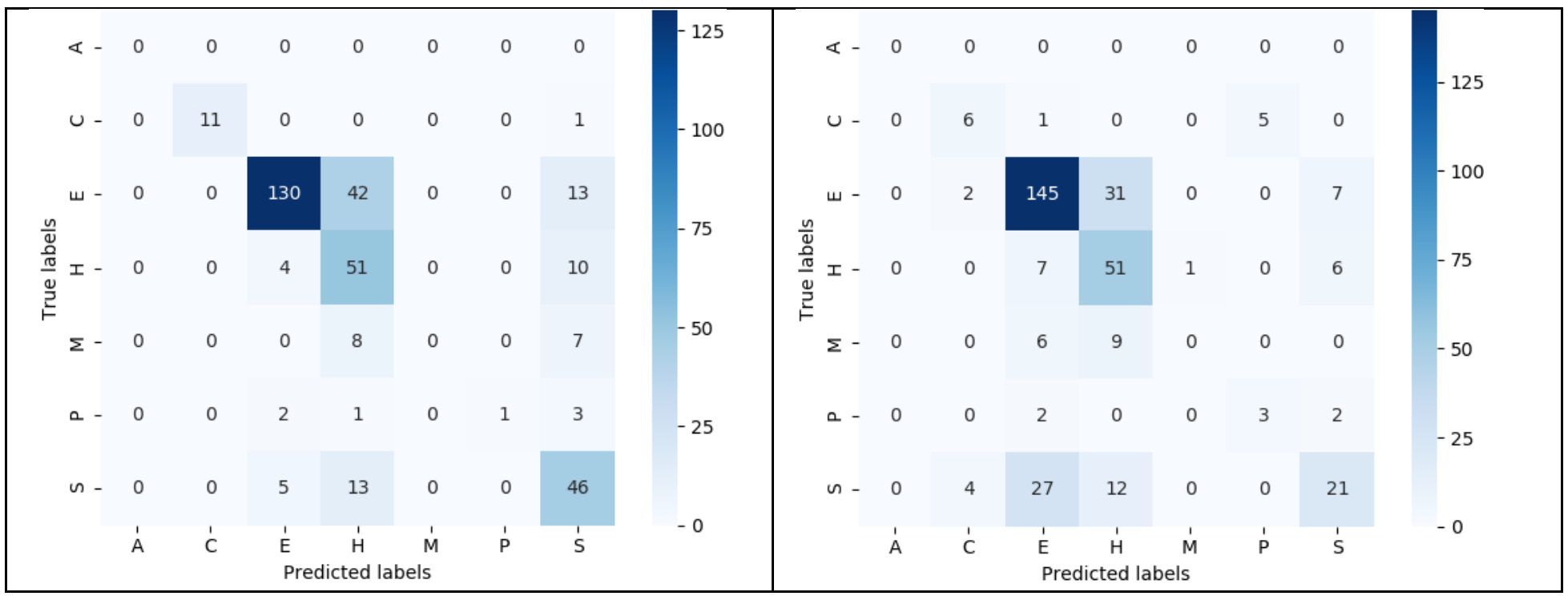

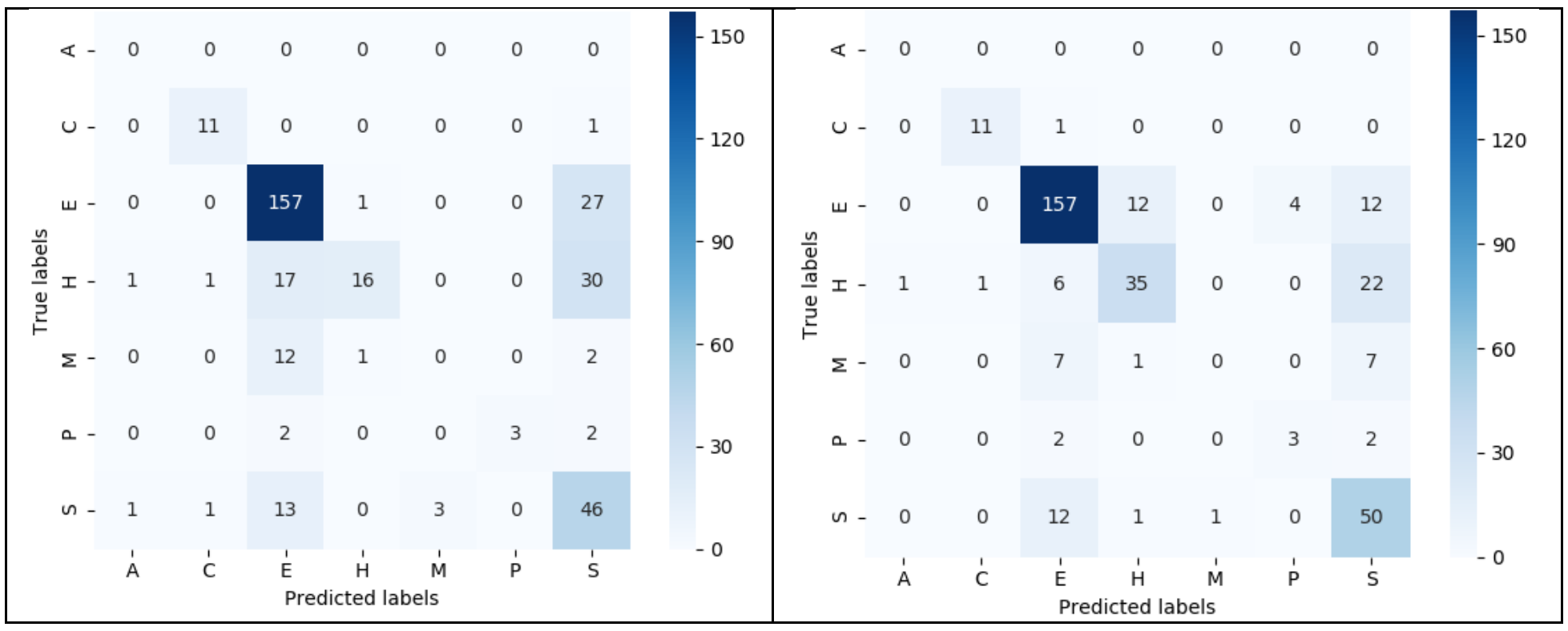

**Fig. 7.** Classification of records from 1st quarter 2019 based on data from: 2018, 2017-2018, 2016-2018 and 2015-2016 (left to right, top-bottom). Numbers show the number of records.

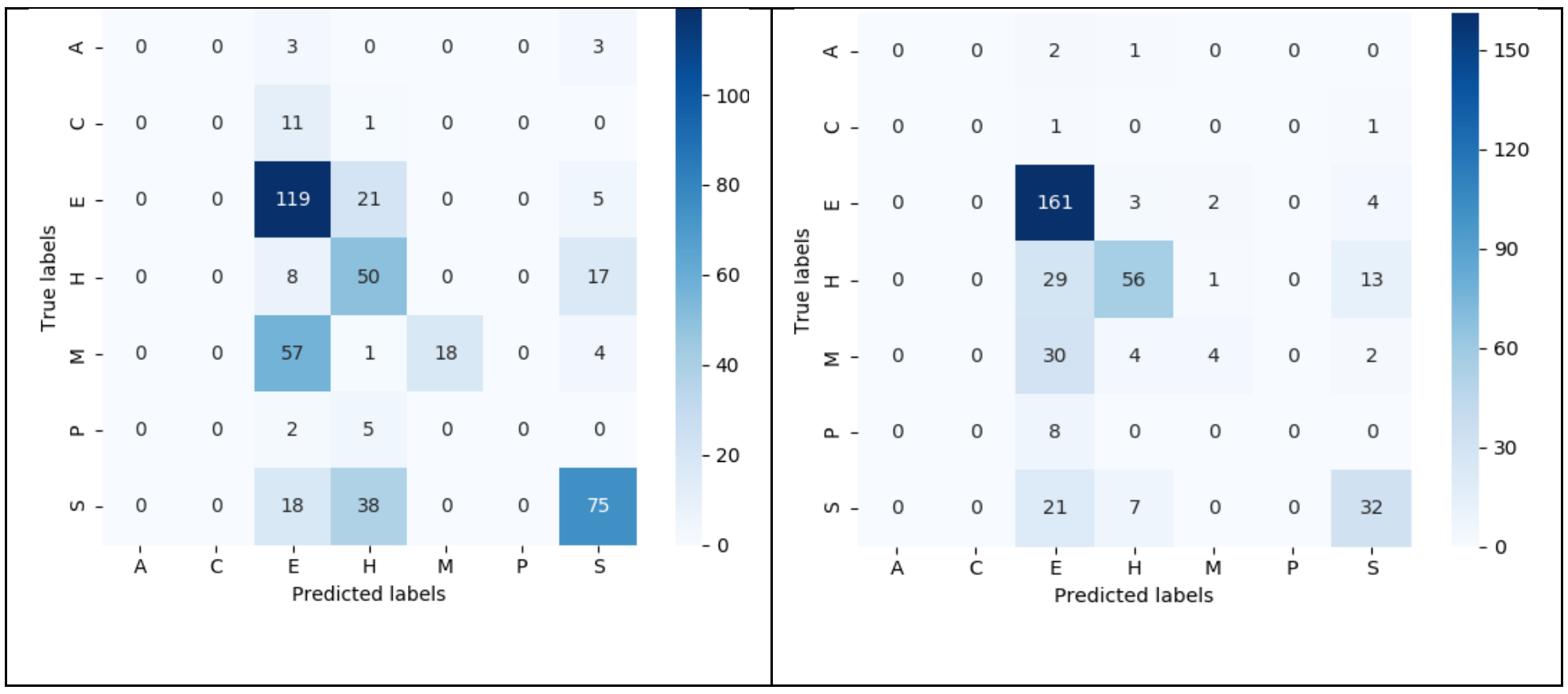

**Fig. 8.** Classification of records from 2015 based on data from 2014 (left) and 2016 based on data from 2015 (right). Numbers show the number of records.

## 4.2 Discussion

As we have shown in the previous section, quality of classification results can be summarized as average. We were able to achieve 70-80% of correct labeling for the most populated classes of devices. In some cases the classification falls below 50%. Using more training data, i.e. going back in time does not always improve the classification quality, on the contrary – in most cases it reduces it.

To summarize the classification results for the whole period of 2011-2019(Q1): **Fig. 9** shows the values of *F1* measure weighted by support (the number of true instances for each label). Because of the weighing, this shows the quality of classification for all classes. As shown, the balanced *F1* score varies between 50% and 72%.

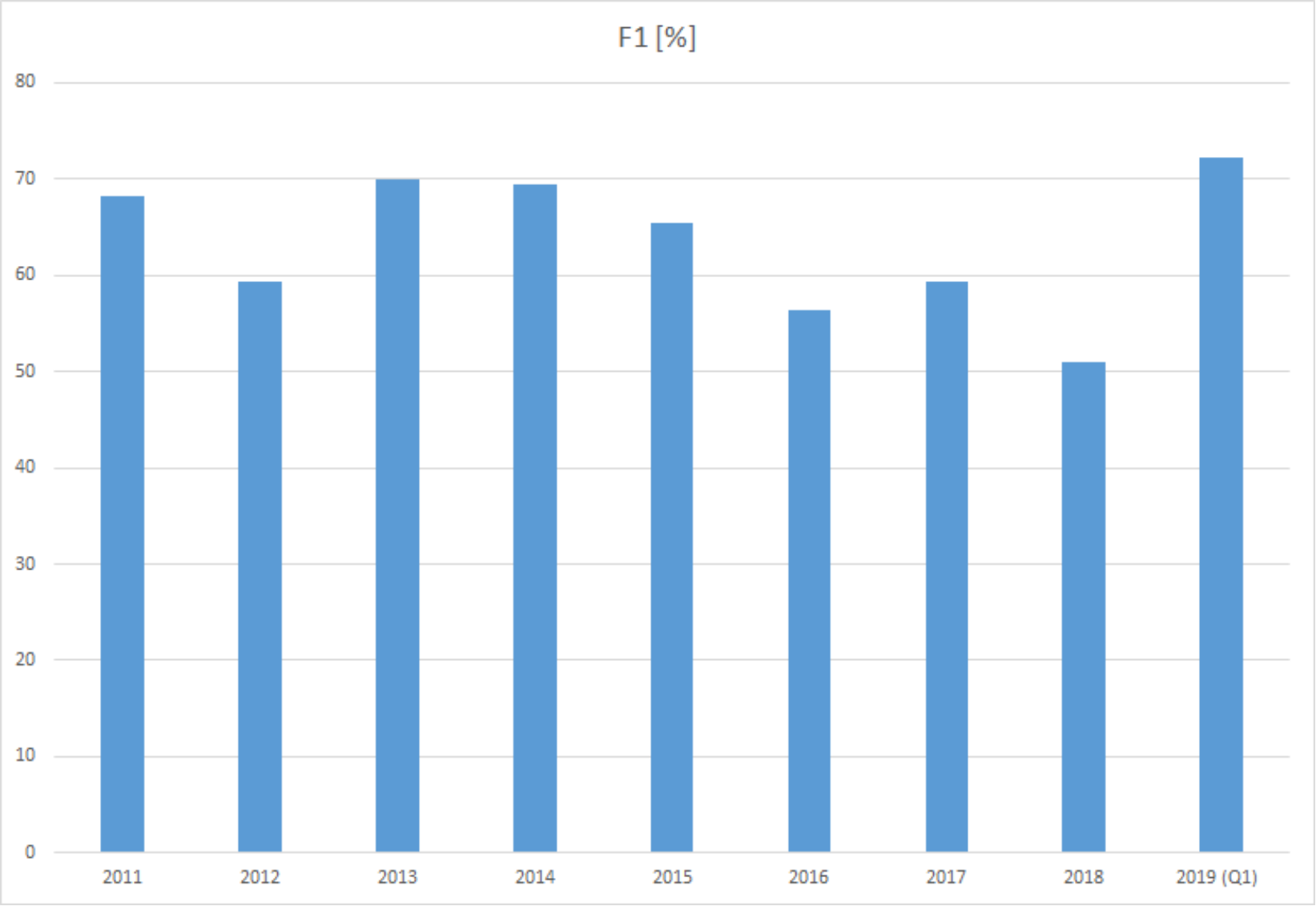

**Fig. 9.** F1 score for records from 2011 to 2019(Q1).

# 5 Summary

We have proposed a classification of IoT device related vulnerability data from the public CVE/NVD database. We have divided vulnerability records into 7 distinct categories: Home and SOHO, SCADA, Enterprise & Networking, Mobile devices, PC devices and other non-home appliances. The hand-classified database samples were used to train a SVM classifier to predict categories of "new" vulnerabilities.

The purpose of the automatic classifier is to predict, and (if possible) in subsequent steps - prevent and mitigate threats resulting from new vulnerabilities. This is not a trivial task to execute by hand given the size of the database and the rate of its growth - when a new vulnerability or exploit is discovered it is often critical to learn its scope by automatics means, as fast as possible.

We have attained classification precision and recall rates of 70-80% for strongly populated categories and of approx. 50% or lower for less numerous categories. This are not ideal results, and in practice they would require further human intervention (verification and possibly reclassification). On the other hand, SVM classifiers have been proved numerous times to be an accurate mechanism for text data classification. The problem in our case lies with the data itself - neither CVE nor CPE contents provide enough specific data for the SVM to discern record categories. We can conclude that the vulnerability ontology should be extended to provide this additional information. Similar conclusions, although not directly related to IoT security, have been drawn by other researchers – e.g. in [36] the authors propose a unified security cybersecurity ontology that incorporates and integrates heterogeneous data and knowledge schemas from different cybersecurity systems, including data about products and product vendors.

Finally, is also worth mentioning that the method used by us is not necessarily limited to CVE database, numerous other on-line vulnerability databases exists [37], which are managed by companies (e.g. Microsoft Security Advisories, TippingPoint Zero Day Initiative, etc.), national CERTs or by professionals' forums (e.g. BugTraq, Exploit-DB, and others). Information from various sources can be integrated and categorized by the method we proposed in this paper. This should increase the precision of the classification and is a topic of our further research.